\documentstyle[aps,multicol,epsf]{revtex}
\begin{document}
\draft
\title{Singularities and Avalanches in Interface Growth
with Quenched Disorder}

\author{Hern\'an A. Makse}

\address{Center for Polymer Studies and Dept. of Physics,
 Boston University, Boston, Massachusetts 02215 }

\maketitle

\begin{abstract}
A simple model for an interface moving in a disordered medium is
presented. The model exhibits a transition between the two
universality classes of interface growth phenomena.
Using this model, it is shown that the application of constraints
to the local slopes of the interface produces avalanches of growth,
that become relevant in the vicinity of the depinning transition.
The study of these avalanches reveals a singular behavior that
explains a recently observed singularity in the equation of
motion of the interface.
\end{abstract}


\begin{multicols}{2}
\narrowtext

  The problem  of interface motion
in  disordered media has attracted
considerable
attention recently.
By means of numerical,  analytical,
and experimental studies \cite{review}
it has been observed that scaling theory can be
used as the framework to understand the
dynamical properties of the interface.
In a typical realization of the problem a $d$-dimensional
interface
moves in a  $(d+1)$-dimensional disordered medium,
driven by an external force $F$.
The interface is assumed to be  oriented
along the longitudinal $\vec{x}$-direction,
and is specified by the  height $y(\vec{x},t)$.
For small forces the interface is pinned by the random quenched
impurities of the medium, and the average velocity is zero. Above
a critical value $F_c$, the
external force overcomes the effect of the
impurities, and the interface
moves with a finite velocity.
Near the threshold, the velocity scales as
$v \sim f^{\theta}$,
where $f \equiv F/F_c-1 $ is the  reduced force, and
$\theta$  the  velocity exponent.

The roughness exponent $\alpha$ is defined by the relation $W(L)
\sim L^\alpha$, where  $W$ is the  width of the interface and $L$
is the system size. The exponent $\alpha$
characterizes the different universality
classes of interface growth phenomena. The universality classes
can also be classified \cite{amaral}
according to the behavior of the coefficient $\lambda$
of a nonlinear term of the
form
$ \lambda (\nabla y)^2$
in the equation of motion of the interface.
One of the universality classes is
described by an equation of the Kardar-Parisi-Zhang (KPZ) type
\cite{kpz} with a
quenched random field $\eta(x,y)$
\begin{equation}
\partial_t y(x,t) =  \nabla^2 y+ \lambda (\nabla y)^2 +
\eta(x,y) + F,
\label{qkpz}
\end{equation}
with the coefficient $\lambda$ diverging at
the depinning transition as
\begin{equation}
\label{lambda}
\lambda(f) \sim f^{-\phi},
\end{equation}
where $\phi \simeq 0.64$ \cite{amaral}.
The roughness exponent at the depinning transition in $(1+1)$-dimensions
is $\alpha = 0.63$, and can be  obtained by a mapping to
directed percolation \cite{tang,sergey}.
We refer to this universality class as directed percolation depinning (DPD).

A second universality class is described, at the depinning transition, by
an equation of the  Edwards-Wilkinson type \cite{ew} with quenched disorder
\begin{equation}
\partial_t y(x,t) =  \nabla^2 y + \eta(x,y) + F.
\label{qew}
\end{equation}
 Models with $\lambda=0$ (for any force), or $\lambda \to 0$ (when $F \to
F_c^+$, corresponding to a negative exponent $\phi$ in Eq. (\ref{lambda})),
belong to the
universality class of Eq. (\ref{qew}).
Analytical \cite{theory} and numerical \cite{lech,dong}
studies
in $(1+1)$-dimensions
yield a roughness exponent $\alpha
\simeq 1.23-1.25$.
This universality class is referred to  as
quenched Edwards-Wilkinson (QEW).

In this paper we present a simple model for an elastic interface
that exhibits a transition between
the two universality classes of interface growth in a disordered
medium.
The DPD result is obtained when lateral or
longitudinal fluctuations of the interface
height are allowed and the growth rule is
restricted by a generalized  solid-on-solid (SOS)
condition.
On the other hand, when the constraint
is absent, the QEW universality class is obtained.
The generalized SOS condition is a
universal feature appearing in all the models of the DPD universality
class.
It is analogous to the  restricted
SOS condition $(|y(x\pm1,t) - y(x,t)| \leq 1)$
presents in simple models of growth with
{\it time-dependent} noise \cite{kk}.
In these type of models, the constraint is commonly associated with
lateral propagation of growth that generates
nonlinearities in the equation of motion.
Moreover, it is shown that, in the presence of {\it quenched} disorder,
the  generalized SOS condition also generates
{\it avalanches of growth}.
These avalanches are  irreversible growth events that
become relevant near the depinning transition.
We show that they present a singular behavior that
is responsible for
the divergence in the coefficient $\lambda$.

Consider a one-dimensional elastic
interface moving in a two-dimensional disordered medium
of lateral size $L$.
A discrete model for such an interface
is defined  in the square lattice by the height values $\{ y_k \}_{k=1,
\cdots , L}$. The interface moves under the influence of an external
force $\vec{F} = (0,F)$ and fluctuates in the {\it longitudinal} or
$x$-direction, and {\it transversal} or $y$-direction.
The strength of these  fluctuations is
controlled by the elastic constants $\nu_x$ and $\nu_y$, respectively.
For the $k$-th column and at a given
time, the velocity $\vec{v} = (v_x,v_y)$ is calculated according to
\begin{equation}
\begin{array}{lcl}
v_y & = & \nu_y \left(y_{k+1} + y_{k-1} - 2 y_k \right) + \eta(k, y_k) + F\\
v_x& =& - 2 \nu_x + \eta(k, y_k).
\end{array}
\label{vxvy}
\end{equation}
The term $\eta(k,y_k)$ represents the strength of the
random force,
and mimics
the effect of quenched disorder. It is a random field
with a uniform
probability distribution between $[ -\delta, \delta]$, and  represents
a repulsive force for positive values and an attractive or
pinning force
for negative values.
A site at the interface moves forward or laterally only when the
respective velocity is positive.
Therefore, the $k$-th column and its nearest neighbors are
updated in the following way

\begin{equation}
\label{update}
\begin{array}{lcll}
y_k& =& y_k +1 & ~~~\mbox{if ~~ $v_y >  0$} \\
y_{k+1}& =& y_k & ~~~\mbox{if ~~ $v_x > 0$} \\
y_{k-1}& =& y_k & ~~~\mbox{if ~~ $v_x > 0$.}
 \end{array}
\end{equation}

The first equations in (\ref{vxvy}) and (\ref{update}),
correspond to a model
introduced to
study an elastic interface
described by the QEW equation \cite{lech}, and they are
equivalent to the discretization of Eq. (\ref{qew}).
The last   equations in (\ref{vxvy}) and (\ref{update}) are  the
simplest generalization of the model
to include longitudinal motions \cite{rsos}.
A lateral motion to a nearest neighbor column is
allowed if the neighboring  column is smaller than the column
considered.
After the lateral motion, the interface
becomes a multi-valued function of the longitudinal coordinate $x$ (see
Fig.\ref{sketch}$(a)$).
The  generalized SOS condition is then applied in order to
transform the interface into a single-valued function,
by defining the interface with the highest value of the height.
This growth process can be  thought as a
{\it transversal  avalanche of growth}
in the  neighboring column, as shown in Fig.\ref{sketch}$(b)$.
This growth event
occurs regardless the value of the noise in the neighboring column.

We perform numerical simulations on a square lattice with
$L=1024$. Helical boundary conditions $y_{L} = y_1 + m ~ L$, where
$m\equiv\langle\nabla y\rangle$ is the
average external tilt, are imposed on the interface in order to study
the interface velocity as a function of the tilt \cite{tilt}.
Figure \ref{parabolas}
shows the tilt-dependence of the velocity of the  interface
for the value $\nu_x=0.2$. In the following, we use the values $\nu_y=1.0$
and $\delta = 3.0$ since our results do not depend of these parameters.
We see that the parabolas
become steeper when $F \to F_c^+$ as in Eq. (\ref{lambda}),
indicating that the model belongs to the DPD
universality class.

The model presents a transition at a critical value
$\nu_{x_c}$.
If the value of $\nu_x$ is increased such that
$\nu_x \geq \nu_{x_c} =
\delta/2$, then the QEW result is found. Specifically, we
find that $v$ is independent of $m$, indicating that
$\lambda = 0$ for any force.
The fact that for $\nu_x > \delta/2$ the $x$-component of
the velocity $v_x$  is always negative explains this transition:
longitudinal motions cannot occur and
one  recovers the model of
Ref.\cite{lech} with the corresponding behavior of  $\lambda$.

The different parabolas obtained for a
given value of $\nu_x <\nu_{x_c} $,
can be rescaled
using the  scaling ansatz \cite{amaral}
\begin{equation}
\label{ansatz}
v(m,f) \sim f^\theta ~ g(m^2 / f^{\theta+\phi}),
\end{equation}
where $g$ is a universal scaling function.
Support for (\ref{ansatz})
is provided by the data collapse shown
in Fig. \ref{data_collapse}, where
we replot the data of Fig. \ref{parabolas} and the
data obtained for $\nu_x=0.6$,
$\nu_x=1.0$, and  $\nu_x=1.4$  \cite{exponents}.
The scaling function $g$ becomes flatter as
$\nu_x \to \nu_{x_c}^-$, indicating that the
prefactor of $\lambda$ goes to zero, even though, the
singularity associated to $f$ is still present as long
as $\nu_x < \nu_{x_c}$.

In the following, we
argue that the divergence in $\lambda$ is explained by the singular
behavior of the size of
the avalanches of growth.

According to (\ref{update}), the interface can advance in two
independent ways. One way is via a transversal motion (if $v_y >0$),
and the other is via a longitudinal motion (if $v_x > 0$) plus the
transversal
avalanche in the neighboring column.
In order to
determine the relevant growth mechanism
near the
depinning transition, we study the mean value
of the number  of
longitudinal and
transversal motions per unit time $n_x$ and $n_y$, respectively,
as a function of the force. For $\nu_x = 0.2$, we find
\begin{equation}
\label{nxny}
\begin{array}{l}
n_x(f) \sim f^{\gamma_x} \\
n_y(f) \sim f^{\gamma_y} ,
\end{array}
\end{equation}
with $\gamma_x \simeq 0.60 $ and $ \gamma_y \simeq 0.78$.  Both
quantities go to zero at $F_c$, since the velocity vanishes at the
depinning transition. However,
the ratio
 $n_x/n_y \sim f^{-0.18}$ diverges at $F_c$, indicating the
{\it relevance} of  longitudinal fluctuations for the motion of the interface
at the depinning transition. This fact can be explained as follows.
For forces close to $F_c$ the velocity is almost
zero and  the interface
moves in a very irregular way, jumping from one metastable
pinning configuration to another. In this ``jerky''
motion the interface takes advantage of longitudinal motions,
rather than transversal ones, to surround and
overcome the impurities.
Thus, {\it avalanche events become relevant
for the motion
of the interface only in the vicinity of the
depinning transition}.

We also
study the mean value of the size
of the avalanches produced by the generalized
SOS condition per unit time,
$\langle s \rangle$,
as a function of the tilt and for different forces.
Figure \ref{avalanches} shows
the results. As it turns out, $\langle s \rangle$,
as well as the
velocity, has a parabolic dependence on $m$. These parabolas become steeper as
$F \to F_c^+$, and we can fit
$\langle s \rangle$
to
\begin{equation}
\label{lambdas}
\langle s \rangle = s_o + \lambda_s ~ m^2
\end{equation}
with $\lambda_s \sim f^{-\phi}$,
and $\phi \simeq 0.64$ the same exponent
as in Eq. (\ref{lambda}).
The parabolas can also be rescaled using the scaling ansatz of
Eq. (\ref{ansatz}) with the same value of $\theta$ used for the velocity
curves.

As shown in Fig. \ref{sketch}$(c)$, the size of an avalanche is larger
for the tilted interface than for the untilted one. This explains the
increase of $\langle s \rangle$ with the tilt,
exemplified in Fig. (\ref{avalanches}) for a given fixed force. Moreover,
since the relative occurrence of lateral motions and avalanches
is larger near the depinning transition
than away from it, the same external tilt will cause a larger
increase in the average
$\langle s \rangle$ near $F_c$ than for $F \gg F_c$.
Thus, the coefficient $\lambda_s$, that measures the variation
experienced by $\langle s \rangle$
due to a change in the average tilt of the interface, increases its
value as $F \to F_c$ and the parabolas become steeper.

Notice that the relevance of longitudinal motions and
avalanches of growth near $F_c$
implies that the velocity is  determined
by the size
of the avalanches,
$v \propto \langle s \rangle$, so that $\lambda \propto \lambda_s$.
Thus, the singularity of the coefficient
$\lambda$ can be explained by the same divergence observed in $\lambda_s$.
The  motion of the interface near
the depinning transition is entirely
dominated by the avalanches of growth produced by the generalized
SOS condition.

The same behavior can be predicted for the other models
of the DPD universality class. All these models share a
constraint in the growth rule of the interface height that generates
avalanches
analogous to the ones in our model.
In the model of Ref. \cite{tang} a slope constraint is applied to the
interface that implies a readjustment of the height regardless of the
value of the noise.
The so-called erosion of
overhangs
in the model of Ref. \cite{sergey} corresponds to our generalized
SOS condition. And
the model of Ref. \cite{sneppen} presents a  restricted
SOS condition that produces
avalanches of growth (see Ref. \cite{tang2}).

Other variants of the model (\ref{vxvy}) are also studied. First, we
study a model  that  includes lateral motions
but removes the generalized SOS condition.
The model can be better understood as a set of $L$-beads
fluctuating in both directions
and interacting through elastic springs \cite{definition}.
We calculate the
velocity for this model as a function of
the average tilt and for different forces.
We find  a small value of $\lambda$ for $F \gg F_c$ and
$\lambda \to 0$ when $F \to F_c^+$.
Therefore, the
model falls in the QEW
universality class. The nonlinearity observed in this model
is of kinematic
origin $(\lambda \propto v)$ \cite{kpz},
in contrast to the behavior of $\lambda$ in
the DPD universality class, for which
$\lambda$ is coupled to the external force and diverges at $F_c$.
These results suggest that lateral motions alone
generate nonlinearities only when the motion of the interface is so
fast that the noise can be regarded as a time-dependent noise . At the
depinning
transition, where the quenched disorder is relevant, the
QEW result is recovered.
These results confirm
that the generalized SOS condition is the origin
of the singularity in $\lambda$.

We also study other mechanisms, which are believed
to generate nonlinearities in the
equation of motion \cite{tkd}: $(i)$  an external fixed force acting in
the normal direction of the interface, analogous to the Lorentz force
found in the motion of
flux lines in type II-superconductors \cite{chao}; and
$(ii)$ an anisotropic random medium characterized by an anisotropic
random force \cite{tkd}. In order to check this
hypothesis, we simulate the model
that allows for longitudinal motions but
without the  generalized SOS condition \cite{why}.
In addition, we generalize the definition of the velocity vector by
including: $(i)$
an external force, with fixed magnitude $F$, acting in the direction
of the local normal vector $\hat{n}$ of the interface:
$\vec{F} \equiv (F_x, F_y)= F \hat{n}$; and $(ii)$
we model the effect of the anisotropy in
the medium by introducing a noise vector
$\vec{\eta} \equiv (\eta_x,\eta_y)$, where $\eta_x$ and $\eta_y$
are independent random fields with amplitudes
$\delta_x$ and  $\delta_y$ respectively ($\delta_x \neq \delta_y$ for an
anisotropic medium).
We find that the model does not belong to the DPD universality class.
Under the limitations of our model, we find that the anisotropy of the medium
does not
generate the divergence in   $\lambda$.

In conclusion we present a simple model for an  elastic
interface moving in a quenched disordered medium, which
captures the relevant features of the two universality classes.
The origin of the singular behavior in the equation of motion of the DPD
universality class is explained by a simple microscopic constraint
imposed to the growth of the interface that generates irreversible
avalanches of growth.

	I wish to thank S. Tomassone for many valuable
contributions to this paper.
I also thank L.  Amaral, A. Barab\'asi,
R. Cuerno, S. Harrington, D. Langtry,
K.  Lauritsen, P. Rey, R. Sadr-Lahijany, and H. E.
Stanley for useful suggestions and discussions.
The Center for Polymer Studies is
supported by the National Science Foundation.

\begin{figure}
\narrowtext
\caption{ Sketch  of the effect of lateral
motions of a site at the interface: $(a)$-$(b)$ for an untilted
interface, and $(c)$ for a tilted interface with $m=1$ as defined in the text.
A lateral motion $(a)$ produces
an effective
avalanche $(b)$
in the nearest neighbor column due to the  generalized SOS condition.
The size of the avalanche, $s$,
 is larger for the tilted interface than for the untilted
one, as can be seen comparing $(b)$ and $(c)$. }
\label{sketch}
\end{figure}

\begin{figure}
\narrowtext
\caption{Plot of the velocity as a function of the average tilt
of the interface for
values of the reduced force from $f=0.19$ (bottom) to $f=1.38$
(top). The parameters are $\nu_x=0.2$, $\nu_y=1.0$ and $\delta=3.0$.
Results are averaged over $70$
independent realizations of the disorder.
The ``closing'' of the parabolas shows that a diverging nonlinear
coefficient is present in the equation of motion.}
\label{parabolas}
\end{figure}

\begin{figure}
\narrowtext
\caption{Data collapse of the data of Fig.~2 ($\nu_x=0.2$)
and the data corresponding to $\nu_x=0.6$,
$\nu_x=1.0$ and $\nu_x=1.4$ (shown from bottom to top, respectively),
plotted according
to the scaling relation of Eq.(6). The scaling function becomes flatter as
$\nu_x \to \nu_{x_c}^-=\delta/2$, indicating the transition
to the QEW universality
class. Each set of curves is shifted for clarity.}
\label{data_collapse}
\end{figure}

\begin{figure}
\narrowtext
\caption{Plot of the average size of the avalanches
produced by the generalized SOS condition as a function of the tilt,
for the same forces and parameters as in Fig.~2. }
\label{avalanches}
\end{figure}

\end{multicols}


\begin{references}
\bibitem{rsos}
The first term that appears in the $x$-component of the  velocity in
Eq. (\ref{vxvy})
corresponds to the elastic force that
opposes the lateral motion.
\bibitem{tilt}
According to (\ref{qkpz}), for a tilted interface with tilt $m$, the
average velocity becomes
$v(m) = v_o + \lambda ~ m^2$,
where $v_o$ is the  velocity of the untilted interface \cite{amaral,krug}.
Thus, one can gain
information on the presence and
magnitude of the  nonlinear coefficient $\lambda$,
by monitoring the velocity
of the interface  as a function of
the average tilt, and fitting to a
parabola the obtained curve.
\bibitem{exponents}
We use $\phi \simeq 0.64$ for all values of $\nu_x$ in the data collapse
of Fig. \ref{data_collapse}. For the velocity
exponent we use $\theta \simeq 0.64$ for $\nu_x=0.2$.
However, we find
somewhat smaller values of $\theta$ as $\nu_x$ approaches $1.5$ ($\theta
\simeq  0.40$ for $\nu_x=1.4$). A possible sources for this deviation is
finite size effects.
\bibitem{definition}
The position of the  $k$-th bead on the
interface is defined by the vector
$\vec{r}_k = (x_k,y_k)$.
Without the  generalized SOS
condition,
an elastic term
$ \nu_x (x_{k+1} + x_{k-1} - 2 x_k) $,
like the one defined  for $v_y$ in Eq. (\ref{vxvy}),
can be introduced in  the $x$-component of the velocity.
The dynamics of the
interface is now  defined by
$y_k=y_k+1$ if $v_y>0$,
$x_{k}=x_k+1$ if $v_x>0$, and
$x_{k}=x_k-1$  if  $v_x<0$.
\bibitem{why}
We choose this model because we want to test the hypothesis
independently of the  generalized SOS condition.
Since the model allows for longitudinal motions, it does not rule out the
possibility of a diverging $\lambda$.
\end{references}
\end{document}